\documentclass[aps,prl,twocolumn,groupedaddress,floatfix,showpacs]{revtex4}

\usepackage{graphicx}

\begin{document}

\begin{widetext}
\noindent\textbf{Preprint of:}\\
Alexis I. Bishop, Timo A. Nieminen, Norman R. Heckenberg,
and Halina Rubinsztein-Dunlop\\
``Optical microrheology using rotating laser-trapped particles''\\
\textit{Physical Review Letters} \textbf{92}(19), 198104 (2004)
\end{widetext}

\title{Optical microrheology using rotating laser-trapped particles}

\author{Alexis I. Bishop}
\email[]{A.I.Bishop@hw.ac.uk}
\altaffiliation{Department of Physics, Heriot-Watt University,
Riccarton, Edinburgh, EH14 4AS, Scotland, United Kingdom}
\author{Timo A. Nieminen}
\author{Norman R. Heckenberg}
\author{Halina Rubinsztein-Dunlop}

\affiliation{Centre for Biophotonics and Laser Science, Department of Physics,
The University of Queensland, Brisbane QLD 4072, Australia}

\begin{abstract}
We demonstrate an optical system that can apply and
accurately measure the torque exerted by the trapping beam on a
rotating birefringent probe particle. This allows the viscosity and
surface effects within liquid media to be measured quantitatively on
a micron-size scale using a trapped rotating spherical
probe particle. We use the
system to measure the viscosity inside a prototype cellular structure.
\end{abstract}

\maketitle 


It is well known that light can transport and transfer angular momentum
as well as linear momentum~\cite{beth1936}. In recent years, this has been
widely applied to rotate microparticles in optical tweezers. Such rotational
micromanipulation has been achieved using absorbing
particles~\cite{friese1996pra,friese1998ol}, birefringent
particles~\cite{friese1998nature,higurashi1999}, specially fabricated
particles~\cite{higurashi1994,galajda2001}, and non-spherical particles
using linearly polarized
light~\cite{bayoudh2003,bonin2002,santamato2002,cheng2002,bishop2003,%
galajda2003}
or non-axisymmetric beams~\cite{paterson2001,oneil2002b}. Notably, particles
rotated in rotationally symmetric beams~\cite{friese1998nature,bishop2003}
spin at a constant speed, indicating that the optical torque driving the
particle is balanced by viscous drag due to the surrounding medium. It has
been noted that the optical torque can be measured
optically~\cite{nieminen2001jmo},
allowing direct independent measurement of the drag torque.

For a pure
Gaussian beam that is circularly polarised, each photon also has an
angular momentum of $\pm\hbar$ about the beam axis, equivalent to
an angular momentum flux of $\pm P/\omega$ for the beam, where $P$ is the
power of an incident beam, and $\omega$ is the angular frequency of the
optical field,
which can be transferred to an
object that absorbs the photon, or changes the polarization
state~\cite{beth1936}. The
reaction torque on a transparent object that changes the degree of circular
polarization of the incident light is given by~\cite{nieminen2001jmo}
\begin{equation}
\tau_R = \Delta\sigma P/\omega
\label{eqn1}
\end{equation}
where $\Delta\sigma$ is the change in circular polarisation.
Thus, by monitoring the change in circular polarization
of light passing through an object, the reaction torque on the object
is found. We exploit this simple principle as the basis for an optically
driven viscometer. The viscosity of a liquid can be determined by
measuring the torque
required to rotate a sphere immersed in the liquid at a constant angular
velocity---a concept we have implemented at the micrometer scale using
a system based on optical tweezers. We have rotated birefringent spheres
of synthetically-grown vaterite that were trapped three-dimensionally
in a circularly polarized optical tweezers trap and measured the
frequency of rotation, and the polarization change of light passing
through the particle, to determine the viscosity of the surrounding fluid.

At least four distinct techniques have been used previously for microrheology:
tracking the diffusion of tracer particles~\cite{chen2003f,mukhopadhyay2001},
tracking the rotational diffusion of disk-shaped tracer
particles~\cite{cheng2003}, measurement of correlations in the Brownian
motion of two optically trapped probe particles~\cite{starrs2003b},
and the measurement of viscous drag acting on an optically trapped probe
particle in linear motion~\cite{lugowski2002}. The method we use here---direct
optical measurement of the rotation of a spherical
probe particle in a stationary
position---allows highly localized measurements to be made, since the probe
particle does not move in the surrounding medium, and, compared with
the rotational diffusion measurements by Cheng and Mason using
microdisks~\cite{cheng2003}, the use of spherical probe particles greatly
simplifies the theoretical analysis of the fluid flow. Also, since the probe
particle is rotationally driven by the optical trap, the rotation rate can
be readily controlled.

The expression for the drag torque, $\tau_D$, on a sphere rotating in
a fluid for low Reynolds number flows is well known~\cite{landau1987},
and is given by
\begin{equation}
\tau_D = 8 \pi\mu a^3 \Omega
\label{eqn2}
\end{equation}
where $\mu$ is the viscosity of the fluid, $a$ is the radius of the sphere,
and $\Omega$ is the angular frequency of the rotation. Equating the
expressions for torques $\tau_R$ and $\tau_D$ gives:
\begin{equation}
\mu = \Delta\sigma P / ( 8\pi a^3 \Omega\omega ).
\label{viscosity}
\end{equation}

This equation suggests that the viscosity can be determined using knowledge
of only a few elementary parameters---the power and the polarization
change of the incident beam passing through the particle, and the size
and rotation rate of the particle. The power, polarization state, and
rotation rate can all be measured optically.

\begin{figure}[htb]
\centerline{%
\begin{tabular}{ll}
(a) & (b) \\
\includegraphics[width=0.47\columnwidth]{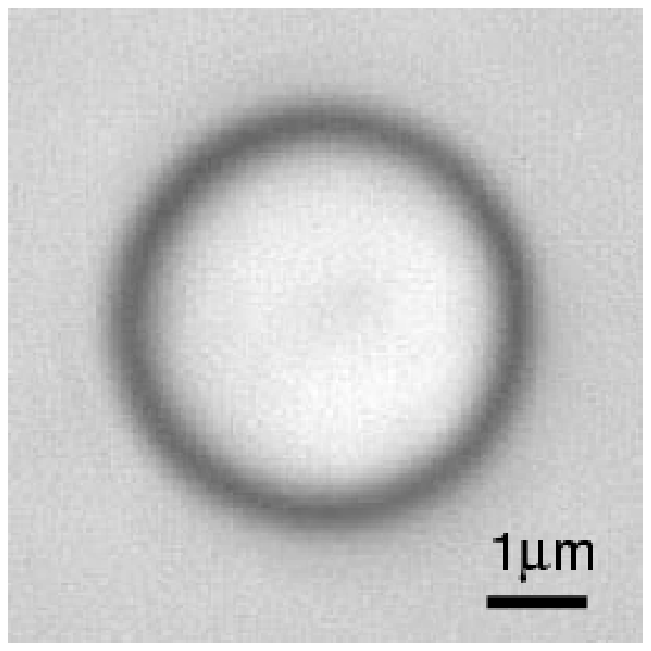} &
\includegraphics[width=0.47\columnwidth]{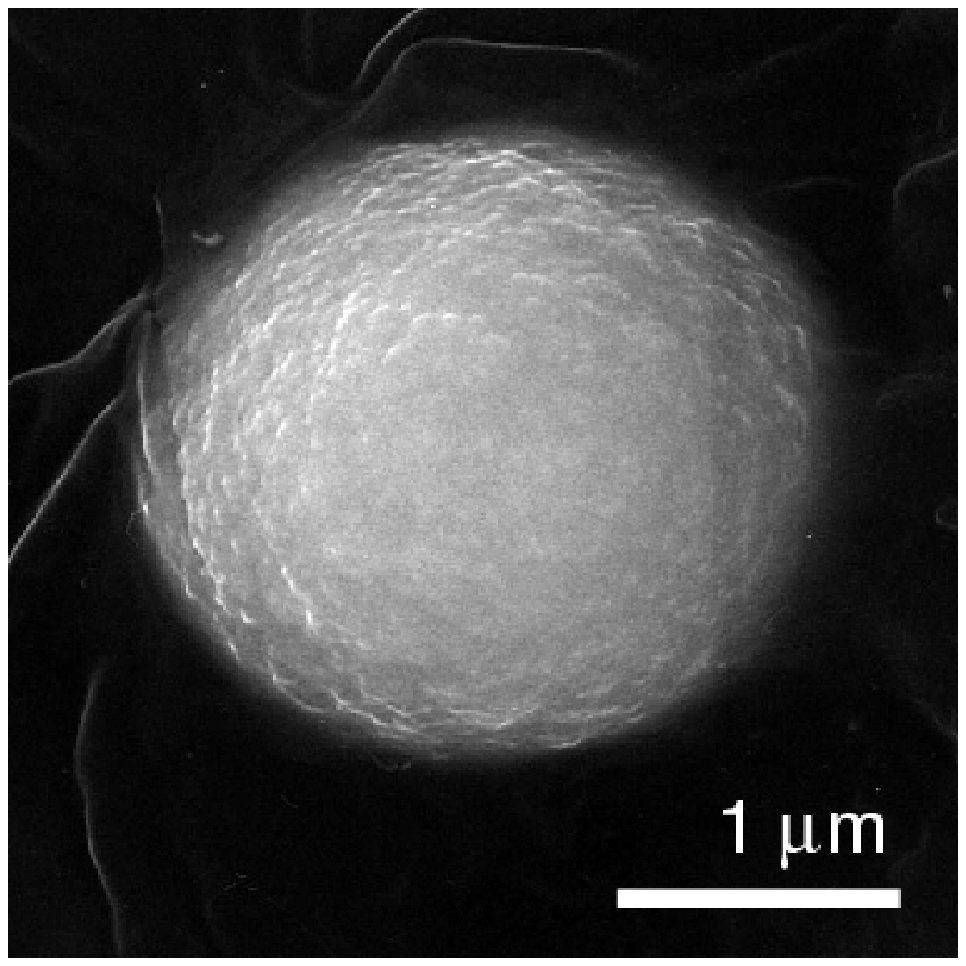}
\end{tabular}
}
\caption{(a) Optical microscope image of a typical vaterite crystal
used for viscosity measurements. (b) Scanning electron microscope
image of a vaterite crystal.}
\label{fig1}
\end{figure}

A significant obstacle to implementing this approach has been obtaining
spherical, transparent, birefringent particles of suitable size for
trapping (approximately 1--10\,$\mu$m in diameter). We have developed a
novel technique to grow nearly perfectly spherical crystals of the
calcium carbonate mineral vaterite, which has similar birefringence
properties to calcite, within the required size regime.
We produce the vaterite crystals by adding 6 drops of 0.1\,M
K$_2$CO$_3$ to a solution of 1.5\,ml of 0.1\.M CaCl$_2$ plus
4 drops of 0.1\,M MgSO$_4$. The solution is strongly agitated by pipetting.
The solution initially appears milky, becoming clearer as the crystals
form within a few minutes of mixing. The typical mean size is found
to be 3\,$\mu$m with a spread of approximately $\pm 1.5\,\mu$m.
Figure~\ref{fig1} shows both an optical image
and an electron micrograph of typical crystals used for viscosity
measurements. Vaterite is a positive uniaxial birefringent material,
with $n_o = 1.55$ and $n_e = 1.65$.

A typical optical tweezers arrangement was used as the basis for the
experiment, which delivered approximately 300\,mW of light at 1064\,nm
into a focal spot with a diameter of around 0.85\,$\mu$m using a $100\times$
oil-immersion objective of high numerical aperture ($\mathrm{NA} = 1.3$).
A $\lambda/4$ plate located immediately before the objective converts
the linearly polarized light from the laser into a circularly polarized
beam which is able to rotate birefringent objects~\cite{friese1998nature}.
Light passing through the probe particle is collected by an oil coupled
condenser lens ($\mathrm{NA} = 1.4$) and then sent to a series of
polarization measuring devices. The degree of circular polarization
remaining in the beam is determined by passing it through a $\lambda/4$
plate with the fast axis orientated at 45$^\circ$ to the axes of a
polarizing beamsplitter cube. Photodiodes monitoring the orthogonal
outputs of the beamsplitter provide a measure of the degree of circular
polarization remaining in the beam, which allows the change in angular
momentum, and hence the torque applied to the particle to be known. The
photodiodes are calibrated such that they provide a measure of the power
at the trap focus. The sum of the two signals gives the total trapping
power, and the difference between the signals gives the degree of
circular polarization of the beam measured in units of power at the
trap focus. The torque applied to the particle is determined by measuring
the difference between the initial degree of circular polarization of
the beam in the absence of the particle and the final
polarization, in accordance with equation~(\ref{eqn1}). The frequency of
rotation of the particle can be accurately determined by monitoring the
transmitted light through a linear polarizer, as suggested by Nieminen
et al.~\cite{nieminen2001jmo}. A small amount of light is diverted from
the main beam and is passed through a polarizing beamsplitter cube,
and the forward direction is monitored using a photodiode. The signal
is modulated at twice the rotation rate of the
particle, and the depth of the modulation is proportional to the
amount of linearly polarized light in the beam. If the particle is
thick enough to act as a $\lambda/2$ plate, then circularly polarized
light passing through the particle is reversed in handedness and the
frequency of the rotation cannot be measured as there is no linear
component to the transmitted light. However, as the particles are
spherical and are of similar diameter to the beam, the polarization at
different radial distances from the rotation axis varies, and there
will almost always be some linear component after transmission.

The calculated viscosity depends on the cube of the radius of the
particle, and thus it is important to determine the size of the
particles accurately. The particles used for viscosity measurements are
in the range of 1.5 to 3.5\,$\mu$m in diameter and consequently to
obtain the viscosity within 10\% requires the diameter to be measured
with an accuracy on the order of 90\,nm, and it was found that
obtaining an accurate measurement of the diameter by direct
visualization was problematic. The method of measurement preferred
by the authors is to put two spheres of nearly identical size in contact,
and measure the distance between centres, which can be found accurately
due to the spherical symmetry. The diameter is inferred using
elementary geometry, and the error in the measurement by this method
is estimated as being $\pm 40$\,nm. The degree of asphericity for typical
particles was estimated to be less than 3\%.

\begin{figure}[htb]
\centerline{\includegraphics[width=\columnwidth]{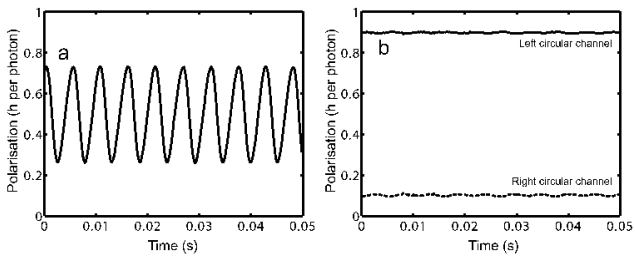}}
\caption{a. Signal recorded by the linear polarization measurement
apparatus during rotation of a vaterite crystal. The frequency of
rotation is $(94.0 \pm 0.7)$\,Hz. b. Signal traces from the circular
polarization measurement apparatus during rotation of a vaterite crystal.
}
\label{fig2}
\end{figure}

The viscosity of water was measured using the system described.
A range of vaterite particles were suspended in distilled water and
trapped three dimensionally in a sealed slide/coverslip cell, with a
depth of 50\,$\mu$m. Particles that are trapped begin to rotate
immediately. However, the rotation may be stopped by aligning the $\lambda/4$
plate located before the objective so that the polarization is made
linear. The size of the particles ranged from 1.5 to 3.5\,$\mu$m in
diameter, and rotation rates up to 400\,Hz (24000\,rpm) were observed
for powers up to 350\,mW at the trap focus. The Reynolds number
of the fluid flow around the rotating spheres is quite low, on the
order of $1\times 10^{-3}$ for rotations up to 1000\,Hz, and hence the
flow is well within the creeping flow regime required by (\ref{eqn2}).
The signals from the circular polarization and linear polarization
monitoring photodiodes were recorded using a 16 bit analogue to
digital converter sampling at 10\,kHz for a period of 5 seconds.
Typical signals recorded during an experiment are reproduced in
figure~\ref{fig2}. The strong uniform modulation of the linear
polarization signal (figure~\ref{fig2}(a)) shows that the particle
rotates very uniformly, taking a large fraction of the angular
momentum carried by the beam. Using a particle of diameter 2.41\,$\mu$m,
the viscosity was found to be $\mu = (9.6\pm0.9)\times 10^{-4}$Pa\,s,
which is in excellent agreement with the established value for the
viscosity of water at 23$^\circ$C of $9.325\times 10^{-4}$Pa\,s~\cite{crc55}.
A series of measurements was made with a range of different sized
particles at a range of rotation rates and trapping powers. The variation
in viscosity over a range of powers below 100\,mW was found to be on the
order of 0.7\% using a single particle. The variation in inferred
viscosity using 11 particles, spanning a range of sizes, was
approximately 4.7\%. The most significant sources of error arise from
the calibration of the trapping power at the focus of the trap (7\%),
the asphericity of the particles (3\%), resulting in a 6\% error in
the viscosity, and the uncertainty in the measurement of the particle
size (1.4\%), which contributes to a 4\% error in the viscosity.
The total error for the viscosity measurement from all error
sources is estimated as being 10\%.

\begin{figure}[htb]
\centerline{\includegraphics[width=0.9\columnwidth]{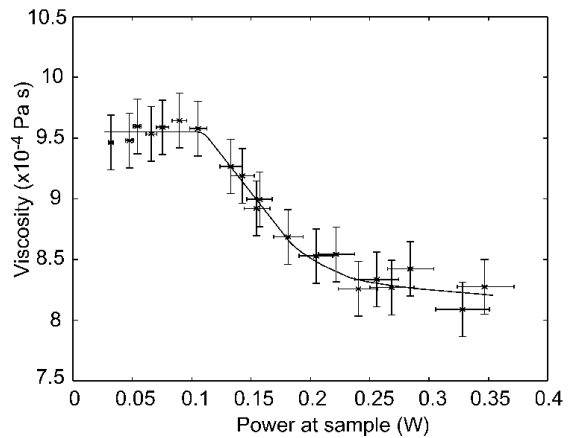}}
\caption{Variation in viscosity with trapping power for a vaterite
crystal of diameter 2.41\,$\mu$m. The viscosity is independent of
power for powers at the trap focus less than approximately 100\,mW.
}
\label{fig3}
\end{figure}

The inferred viscosity was found to be independent of rotation rate,
although for trapping powers above approximately 100\,mW, the viscosity
was observed to decrease with increasing power. This is expected,
due to heating of the particle which results from the absorption of
light through the sample, and also heating of the surrounding liquid
due to absorption of the trapping beam by water, which is on the
order of 10\,K per W at the focus~\cite{liu1996b}. Figure~\ref{fig3}
demonstrates the behaviour of the measured viscosity as a function
of the power at the trapping focus. It was observed that higher trapping
powers
could be used with smaller particles before the onset of the decrease
in viscosity, which is consistent with volume absorption effects. The
observed decrease in viscosity with power was initially approximately
linear and was approximately 0.13\% per mW, which is equivalent to a
temperature rise of around 0.06$^\circ$C per mW. The total power
loss for the trapping beam passing through a particle was measured
for a sample of vaterite particles, and was found to be within the
range of 0.6\% to 1.2\%. The reflection loss due to refractive index
mismatches at the particle surfaces is approximately
0.6\%.

\begin{figure}[htb]
\centerline{\includegraphics[width=0.9\columnwidth]{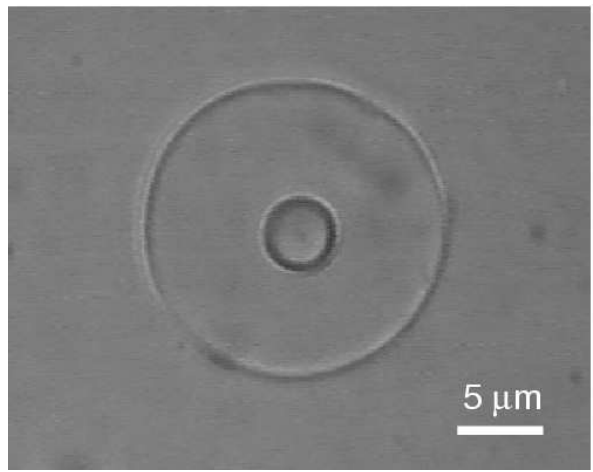}}
\caption{Free vaterite crystal located inside a hexane-filled lipid-walled
vesicle of 16.7\,$\mu$m in diameter.}
\label{fig4}
\end{figure}

We have also demonstrated the ability to measure the viscosity inside
a small confined region, such as a cellular membrane. Hexane filled
vesicles were formed by emulsifying hexane, containing a small amount
of soy lecithin (1\,g/l), with water containing a quantity of vaterite
spheres. Spherical membrane structures were formed with diameters up
to approximately 20\,$\mu$m which occasionally contained single
vaterite crystals. The viscosity inside a 16.7\,$\mu$m diameter vesicle
was measured with a vaterite particle of diameter 3.00\,$\mu$m, using
the same procedure as outlined earlier, and was found to be
$(2.7\pm 0.5)\times 10^{-4}$Pa\,s for powers less than 100\,mW, which
is in good agreement with the established value of viscosity for hexane
at 23$^\circ$C of $3.07\times 10^{-4}$Pa\,s (extrapolated)~\cite{crc83}.
This value is significantly less than that of water
($9.325\times 10^{-4}$Pa\,s at 23$^\circ$C~\cite{crc55}. Figure~\ref{fig4}
shows the hexane filled vesicle with the vaterite probe particle inside.

The viscous drag torque acting on a spherical probe particle of radius
$a$ rotating with an angular frequency $\Omega$
at the center of a sphere of radius $R$ of fluid of viscosity $\mu_1$,
surrounded by a fluid of viscosity $\mu_2$ is~\cite{landau1987}
\begin{equation}
\tau_D = 8 \pi\mu a^3 \Omega R^3 / ( R^3 - a^3 + a^3 \mu_1 / \mu_2 ).
\end{equation}
This can be used to determine the viscosity within a vesicle if the viscosity
of the surrounding fluid is known, or to estimate the error if
equation~(\ref{viscosity}) is used instead. For other geometries, such as
if the probe particle in not in the centre of the vesicle, or is near a plane
interface~\cite{danov1998}, the effect will be more complicated. However, in
most cases the effect of the boundary can be ignored if the probe particle is
at least one diameter away from the interface.
The effect of nearby boundaries was determined experimentally
by measuring the viscous drag
torque at varying distances from a solid interface. For rotation about
an axis parallel to the interface, no change in the drag torque was observed
until the probe particle was within one diameter of the interface.
For rotation about an axis normal to the interface, no change in the drag
torque was observed until the probe particle was very close to the interface.
For the above measurement of the viscosity of hexane, 
the drag in the vesicle is only 0.4\% greater than in pure hexane.
The rotation rate of the vesicle is
$\Omega_R = \Omega / [ 1 + \mu_2 (R^3 - a^3)/ (\mu_1 a^3) ]$~\cite{landau1987},
equal to $\Omega/522$ for the case above. Small particles in the surrounding
fluid near the vesicle were observed to orbit the vesicle at approximately
this rate.

The most significant contribution to the error in these measurements
is the uncertainty in accurately determining the diameter of the
spherical particle. It was not found to be possible to drag a particle
through the membrane, and hence the method of size measurement using
two spheres in contact could not be used. The diameter was instead
directly estimated from an image recorded of the sphere.
The error in determining the diameter is
estimated as being 5\%, which is the main contribution to the total
measurement error of 19\%.

It is feasible to produce smaller vaterite particles, and the use
of these would allow the viscosity to be probed on an even smaller
size scale---potentially probing volumes with a capacity of a cubic
micron. The ability to functionalise the probe particle surface would
enable the selective attachment of the particles to various biological
structures, which would allow the torsional response of these structures
to be investigated quantitatively. We note that the torque and angular
deflection can still be measured accurately in the absence of continuous
rotation. The demonstrated ability to measure viscosity within micron-sized
volumes makes the system developed here of great value for probing the
microrheology of liquid-based materials, such as colloids.

\subsection*{Acknowledgements}

This work was partially supported by the Australian Research Council.
We wish to thank Professor Rane Curl, University of Michigan,
for valuable discussions.


\begin{thebibliography}{23}
\expandafter\ifx\csname natexlab\endcsname\relax\def\natexlab#1{#1}\fi
\expandafter\ifx\csname bibnamefont\endcsname\relax
  \def\bibnamefont#1{#1}\fi
\expandafter\ifx\csname bibfnamefont\endcsname\relax
  \def\bibfnamefont#1{#1}\fi
\expandafter\ifx\csname citenamefont\endcsname\relax
  \def\citenamefont#1{#1}\fi
\expandafter\ifx\csname url\endcsname\relax
  \def\url#1{\texttt{#1}}\fi
\expandafter\ifx\csname urlprefix\endcsname\relax\def\urlprefix{URL }\fi
\providecommand{\bibinfo}[2]{#2}
\providecommand{\eprint}[2][]{\url{#2}}

\bibitem[{\citenamefont{Beth}(1936)}]{beth1936}
\bibinfo{author}{\bibfnamefont{R.~A.} \bibnamefont{Beth}},
  \bibinfo{journal}{Physical Review} \textbf{\bibinfo{volume}{50}},
  \bibinfo{pages}{115} (\bibinfo{year}{1936}).

\bibitem[{\citenamefont{Friese et~al.}(1996)\citenamefont{Friese, Enger,
  Rubinsztein-Dunlop, and Heckenberg}}]{friese1996pra}
\bibinfo{author}{\bibfnamefont{M.~E.~J.} \bibnamefont{Friese}},
  \bibinfo{author}{\bibfnamefont{J.}~\bibnamefont{Enger}},
  \bibinfo{author}{\bibfnamefont{H.}~\bibnamefont{Rubinsztein-Dunlop}},
  \bibnamefont{and} \bibinfo{author}{\bibfnamefont{N.~R.}
  \bibnamefont{Heckenberg}}, \bibinfo{journal}{Physical Review A}
  \textbf{\bibinfo{volume}{54}}, \bibinfo{pages}{1593} (\bibinfo{year}{1996}).

\bibitem[{\citenamefont{Friese et~al.}(1998{\natexlab{a}})\citenamefont{Friese,
  Nieminen, Heckenberg, and Rubinsztein-Dunlop}}]{friese1998ol}
\bibinfo{author}{\bibfnamefont{M.~E.~J.} \bibnamefont{Friese}},
  \bibinfo{author}{\bibfnamefont{T.~A.} \bibnamefont{Nieminen}},
  \bibinfo{author}{\bibfnamefont{N.~R.} \bibnamefont{Heckenberg}},
  \bibnamefont{and}
  \bibinfo{author}{\bibfnamefont{H.}~\bibnamefont{Rubinsztein-Dunlop}},
  \bibinfo{journal}{Opt. Lett.} \textbf{\bibinfo{volume}{23}},
  \bibinfo{pages}{1} (\bibinfo{year}{1998}{\natexlab{a}}).

\bibitem[{\citenamefont{Friese et~al.}(1998{\natexlab{b}})\citenamefont{Friese,
  Nieminen, Heckenberg, and Rubinsztein-Dunlop}}]{friese1998nature}
\bibinfo{author}{\bibfnamefont{M.~E.~J.} \bibnamefont{Friese}},
  \bibinfo{author}{\bibfnamefont{T.~A.} \bibnamefont{Nieminen}},
  \bibinfo{author}{\bibfnamefont{N.~R.} \bibnamefont{Heckenberg}},
  \bibnamefont{and}
  \bibinfo{author}{\bibfnamefont{H.}~\bibnamefont{Rubinsztein-Dunlop}},
  \bibinfo{journal}{Nature} \textbf{\bibinfo{volume}{394}},
  \bibinfo{pages}{348} (\bibinfo{year}{1998}{\natexlab{b}}),
  \bibinfo{note}{erratum in \textit{Nature}, \textbf{395}, 621 (1998)}.

\bibitem[{\citenamefont{Higurashi et~al.}(1999)\citenamefont{Higurashi, Sawada,
  and Ito}}]{higurashi1999}
\bibinfo{author}{\bibfnamefont{E.}~\bibnamefont{Higurashi}},
  \bibinfo{author}{\bibfnamefont{R.}~\bibnamefont{Sawada}}, \bibnamefont{and}
  \bibinfo{author}{\bibfnamefont{T.}~\bibnamefont{Ito}},
  \bibinfo{journal}{Physical Review E} \textbf{\bibinfo{volume}{59}},
  \bibinfo{pages}{3676} (\bibinfo{year}{1999}).

\bibitem[{\citenamefont{Higurashi et~al.}(1994)\citenamefont{Higurashi, Ukita,
  Tanaka, and Ohguchi}}]{higurashi1994}
\bibinfo{author}{\bibfnamefont{E.}~\bibnamefont{Higurashi}},
  \bibinfo{author}{\bibfnamefont{H.}~\bibnamefont{Ukita}},
  \bibinfo{author}{\bibfnamefont{H.}~\bibnamefont{Tanaka}}, \bibnamefont{and}
  \bibinfo{author}{\bibfnamefont{O.}~\bibnamefont{Ohguchi}},
  \bibinfo{journal}{Appl. Phys. Lett.} \textbf{\bibinfo{volume}{64}},
  \bibinfo{pages}{2209} (\bibinfo{year}{1994}).

\bibitem[{\citenamefont{Galajda and Ormos}(2001)}]{galajda2001}
\bibinfo{author}{\bibfnamefont{P.}~\bibnamefont{Galajda}} \bibnamefont{and}
  \bibinfo{author}{\bibfnamefont{P.}~\bibnamefont{Ormos}},
  \bibinfo{journal}{Applied Physics Letters} \textbf{\bibinfo{volume}{78}},
  \bibinfo{pages}{249} (\bibinfo{year}{2001}).

\bibitem[{\citenamefont{Bayoudh et~al.}(2003)\citenamefont{Bayoudh, Nieminen,
  Heckenberg, and Rubinsztein-Dunlop}}]{bayoudh2003}
\bibinfo{author}{\bibfnamefont{S.}~\bibnamefont{Bayoudh}},
  \bibinfo{author}{\bibfnamefont{T.~A.} \bibnamefont{Nieminen}},
  \bibinfo{author}{\bibfnamefont{N.~R.} \bibnamefont{Heckenberg}},
  \bibnamefont{and}
  \bibinfo{author}{\bibfnamefont{H.}~\bibnamefont{Rubinsztein-Dunlop}},
  \bibinfo{journal}{Journal of Modern Optics} \textbf{\bibinfo{volume}{50}},
  \bibinfo{pages}{1581} (\bibinfo{year}{2003}).

\bibitem[{\citenamefont{Bonin et~al.}(2002)\citenamefont{Bonin, Kourmanov, and
  Walker}}]{bonin2002}
\bibinfo{author}{\bibfnamefont{K.~D.} \bibnamefont{Bonin}},
  \bibinfo{author}{\bibfnamefont{B.}~\bibnamefont{Kourmanov}},
  \bibnamefont{and} \bibinfo{author}{\bibfnamefont{T.~G.}
  \bibnamefont{Walker}}, \bibinfo{journal}{Opt. Express}
  \textbf{\bibinfo{volume}{10}}, \bibinfo{pages}{984} (\bibinfo{year}{2002}).

\bibitem[{\citenamefont{Santamato et~al.}(2002)\citenamefont{Santamato, Sasso,
  Piccirillo, and Vella}}]{santamato2002}
\bibinfo{author}{\bibfnamefont{E.}~\bibnamefont{Santamato}},
  \bibinfo{author}{\bibfnamefont{A.}~\bibnamefont{Sasso}},
  \bibinfo{author}{\bibfnamefont{B.}~\bibnamefont{Piccirillo}},
  \bibnamefont{and} \bibinfo{author}{\bibfnamefont{A.}~\bibnamefont{Vella}},
  \bibinfo{journal}{Opt. Express} \textbf{\bibinfo{volume}{10}},
  \bibinfo{pages}{871} (\bibinfo{year}{2002}).

\bibitem[{\citenamefont{Cheng et~al.}(2002)\citenamefont{Cheng, Chaikin, and
  Mason}}]{cheng2002}
\bibinfo{author}{\bibfnamefont{Z.}~\bibnamefont{Cheng}},
  \bibinfo{author}{\bibfnamefont{P.~M.} \bibnamefont{Chaikin}},
  \bibnamefont{and} \bibinfo{author}{\bibfnamefont{T.~G.} \bibnamefont{Mason}},
  \bibinfo{journal}{Phys. Rev. Lett.} \textbf{\bibinfo{volume}{89}},
  \bibinfo{pages}{108303} (\bibinfo{year}{2002}).

\bibitem[{\citenamefont{Bishop et~al.}(2003)\citenamefont{Bishop, Nieminen,
  Heckenberg, and Rubinsztein-Dunlop}}]{bishop2003}
\bibinfo{author}{\bibfnamefont{A.~I.} \bibnamefont{Bishop}},
  \bibinfo{author}{\bibfnamefont{T.~A.} \bibnamefont{Nieminen}},
  \bibinfo{author}{\bibfnamefont{N.~R.} \bibnamefont{Heckenberg}},
  \bibnamefont{and}
  \bibinfo{author}{\bibfnamefont{H.}~\bibnamefont{Rubinsztein-Dunlop}},
  \bibinfo{journal}{Phys. Rev. A} \textbf{\bibinfo{volume}{68}},
  \bibinfo{pages}{033802} (\bibinfo{year}{2003}).

\bibitem[{\citenamefont{Galajda and Ormos}(2003)}]{galajda2003}
\bibinfo{author}{\bibfnamefont{P.}~\bibnamefont{Galajda}} \bibnamefont{and}
  \bibinfo{author}{\bibfnamefont{P.}~\bibnamefont{Ormos}},
  \bibinfo{journal}{Opt. Express} \textbf{\bibinfo{volume}{11}},
  \bibinfo{pages}{446} (\bibinfo{year}{2003}).

\bibitem[{\citenamefont{Paterson et~al.}(2001)\citenamefont{Paterson,
  MacDonald, Arlt, Sibbett, Bryant, and Dholakia}}]{paterson2001}
\bibinfo{author}{\bibfnamefont{L.}~\bibnamefont{Paterson}},
  \bibinfo{author}{\bibfnamefont{M.~P.} \bibnamefont{MacDonald}},
  \bibinfo{author}{\bibfnamefont{J.}~\bibnamefont{Arlt}},
  \bibinfo{author}{\bibfnamefont{W.}~\bibnamefont{Sibbett}},
  \bibinfo{author}{\bibfnamefont{P.~E.} \bibnamefont{Bryant}},
  \bibnamefont{and} \bibinfo{author}{\bibfnamefont{K.}~\bibnamefont{Dholakia}},
  \bibinfo{journal}{Science} \textbf{\bibinfo{volume}{292}},
  \bibinfo{pages}{912} (\bibinfo{year}{2001}).

\bibitem[{\citenamefont{O'Neil and Padgett}(2002)}]{oneil2002b}
\bibinfo{author}{\bibfnamefont{A.~T.} \bibnamefont{O'Neil}} \bibnamefont{and}
  \bibinfo{author}{\bibfnamefont{M.~J.} \bibnamefont{Padgett}},
  \bibinfo{journal}{Opt. Lett.} \textbf{\bibinfo{volume}{27}},
  \bibinfo{pages}{743} (\bibinfo{year}{2002}).

\bibitem[{\citenamefont{Nieminen et~al.}(2001)\citenamefont{Nieminen,
  Heckenberg, and Rubinsztein-Dunlop}}]{nieminen2001jmo}
\bibinfo{author}{\bibfnamefont{T.~A.} \bibnamefont{Nieminen}},
  \bibinfo{author}{\bibfnamefont{N.~R.} \bibnamefont{Heckenberg}},
  \bibnamefont{and}
  \bibinfo{author}{\bibfnamefont{H.}~\bibnamefont{Rubinsztein-Dunlop}},
  \bibinfo{journal}{Journal of Modern Optics} \textbf{\bibinfo{volume}{48}},
  \bibinfo{pages}{405} (\bibinfo{year}{2001}).

\bibitem[{\citenamefont{Chen et~al.}(2003)\citenamefont{Chen, Weeks, Crocker,
  Islam, Verma, Gruber, Levine, Lubensky, and Yodh}}]{chen2003f}
\bibinfo{author}{\bibfnamefont{D.~T.} \bibnamefont{Chen}},
  \bibinfo{author}{\bibfnamefont{E.~R.} \bibnamefont{Weeks}},
  \bibinfo{author}{\bibfnamefont{J.~C.} \bibnamefont{Crocker}},
  \bibinfo{author}{\bibfnamefont{M.~F.} \bibnamefont{Islam}},
  \bibinfo{author}{\bibfnamefont{R.}~\bibnamefont{Verma}},
  \bibinfo{author}{\bibfnamefont{J.}~\bibnamefont{Gruber}},
  \bibinfo{author}{\bibfnamefont{A.~J.} \bibnamefont{Levine}},
  \bibinfo{author}{\bibfnamefont{T.~C.} \bibnamefont{Lubensky}},
  \bibnamefont{and} \bibinfo{author}{\bibfnamefont{A.~G.} \bibnamefont{Yodh}},
  \bibinfo{journal}{Phys. Rev. Lett.} \textbf{\bibinfo{volume}{90}},
  \bibinfo{pages}{108301} (\bibinfo{year}{2003}).

\bibitem[{\citenamefont{Mukhopadhyay and Granick}(2001)}]{mukhopadhyay2001}
\bibinfo{author}{\bibfnamefont{A.}~\bibnamefont{Mukhopadhyay}}
  \bibnamefont{and} \bibinfo{author}{\bibfnamefont{S.}~\bibnamefont{Granick}},
  \bibinfo{journal}{Current Opinion in Colloid and Interface Science}
  \textbf{\bibinfo{volume}{6}}, \bibinfo{pages}{423} (\bibinfo{year}{2001}).

\bibitem[{\citenamefont{Cheng and Mason}(2003)}]{cheng2003}
\bibinfo{author}{\bibfnamefont{Z.}~\bibnamefont{Cheng}} \bibnamefont{and}
  \bibinfo{author}{\bibfnamefont{T.~G.} \bibnamefont{Mason}},
  \bibinfo{journal}{Phys. Rev. Lett.} \textbf{\bibinfo{volume}{90}},
  \bibinfo{pages}{018304} (\bibinfo{year}{2003}).

\bibitem[{\citenamefont{Starrs and Bartlett}(2003)}]{starrs2003b}
\bibinfo{author}{\bibfnamefont{L.}~\bibnamefont{Starrs}} \bibnamefont{and}
  \bibinfo{author}{\bibfnamefont{P.}~\bibnamefont{Bartlett}},
  \bibinfo{journal}{Faraday Discuss.} \textbf{\bibinfo{volume}{123}},
  \bibinfo{pages}{323} (\bibinfo{year}{2003}).

\bibitem[{\citenamefont{Rafa{\l}~{\L}ugowski and Kawata}(2002)}]{lugowski2002}
\bibinfo{author}{\bibfnamefont{B.~K.} \bibnamefont{Rafa{\l}~{\L}ugowski}}
  \bibnamefont{and} \bibinfo{author}{\bibfnamefont{Y.}~\bibnamefont{Kawata}},
  \bibinfo{journal}{Optics Communications} \textbf{\bibinfo{volume}{202}},
  \bibinfo{pages}{1} (\bibinfo{year}{2002}).

\bibitem[{\citenamefont{Landau and Lifshitz}(1987)}]{landau1987}
\bibinfo{author}{\bibfnamefont{L.~D.} \bibnamefont{Landau}} \bibnamefont{and}
  \bibinfo{author}{\bibfnamefont{E.~M.} \bibnamefont{Lifshitz}},
  \emph{\bibinfo{title}{Fluid Mechanics}}, vol.~\bibinfo{volume}{6} of
  \emph{\bibinfo{series}{Course of Theoretical Physics}}
  (\bibinfo{publisher}{Butterworth-Heinemann}, \bibinfo{address}{Oxford},
  \bibinfo{year}{1987}), \bibinfo{edition}{2nd} ed.

\bibitem[{\citenamefont{CRC Handbook}(1974)}]{crc55}
  \emph{\bibinfo{title}{CRC Handbook of Chemistry and Physics, 1974--1975}}
  (\bibinfo{publisher}{CRC Press},
  \bibinfo{year}{1974}), \bibinfo{edition}{55th} ed.

\bibitem[{\citenamefont{Liu et~al.}(1996)\citenamefont{Liu, Sonek, Berns, and
  Tromberg}}]{liu1996b}
\bibinfo{author}{\bibfnamefont{Y.}~\bibnamefont{Liu}},
  \bibinfo{author}{\bibfnamefont{G.~J.} \bibnamefont{Sonek}},
  \bibinfo{author}{\bibfnamefont{M.~W.} \bibnamefont{Berns}}, \bibnamefont{and}
  \bibinfo{author}{\bibfnamefont{B.~J.} \bibnamefont{Tromberg}},
  \bibinfo{journal}{Biophysical Journal} \textbf{\bibinfo{volume}{71}},
  \bibinfo{pages}{2158} (\bibinfo{year}{1996}).

\bibitem[{\citenamefont{CRC Handbook}(2002)}]{crc83}
  \emph{\bibinfo{title}{CRC Handbook of Chemistry and Physics, 2002--2003}}
  (\bibinfo{publisher}{CRC Press},
  \bibinfo{year}{2002}), \bibinfo{edition}{83rd} ed.

\bibitem[{\citenamefont{Danov et~al.}(1998)\citenamefont{Danov, Gurkov,
Raszillier, and Durst}}]{danov1998}
\bibinfo{author}{\bibfnamefont{K.~D.} \bibnamefont{Danov}},
\bibinfo{author}{\bibfnamefont{T.~D.} \bibnamefont{Gurkov}},
\bibinfo{author}{\bibfnamefont{H.} \bibnamefont{Raszillier}}, \bibnamefont{and}
\bibinfo{author}{\bibfnamefont{F.} \bibnamefont{Durst}},
\bibinfo{journal}{Chemical Engineering Science} \textbf{\bibinfo{volume}{53}},
\bibinfo{pages}{3413} (\bibinfo{year}{1998}).

\end{thebibliography}

\end{document}